# Transformer coupling and its modelling for the flux-ramp modulation of rf-SQUIDs

**P. Carniti**[b, a], **L. Cassina**[b, a], **M. Faverzani**[b,a], **E. Ferri**[a, b], **A. Giachero**[b, a], **C. Gotti**[b, a], **M. Maino**[a, b], **A. Nucciotti**[b, a], **G. Pessina**[a, b], **A. Puiu**[b, a]

[a] *INFN – Istituto Nazionale di Fisica Nucleare, sezione di Milano-Bicocca,*
*Piazza della Scienza 3, Milano, 20126-I*

[b] *Università di Milano-Bicocca, Dipartimento di Fisica,*
*Piazza della Scienza 3, Milano, 2016-I*

*E-mail*: Elena.Ferri@mib.infn.it

ABSTRACT: Microwave frequency domain multiplexing is a suitable technique to read out a large number of detector channels using only a few connecting lines. In the HOLMES experiment this is based on inductively coupled rf-SQUIDs (Superconducting QUantum Interference Devices) fed by TES (Transition Edge Sensors). Biasing of the whole rf-SQUID chain is provided with a single transmission line by means of the recently introduced flux-ramp modulation technique, a sawtooth signal which allows signal reconstruction while operating the rf-SQUIDs in open loop condition. Due to the crucial role of the sawtooth signal, it is very important that it does not suffer from ground loop disturbances and EMI. Introducing a transformer between the biasing source and the SQUIDs is very effective in suppressing disturbances. The sawtooth signal has slow and fast components, and the period can vary between a few kHz up to MHz depending on the TES signal and SQUID characteristics. A transformer able to face such a broad range of conditions must have very stringent characteristics and needs to be custom designed. Our solution exploits standard commercial, and inexpensive, transformers for LAN networks used in a suitable combination. A model that allows to take care of the low as well as of the high frequency operating range has been developed.

**Contents**

**1. Introduction**

The use of SQUIDs (Superconducting QUantum Interference Devices) in cryogenic particle detectors allows to implement the read out of large arrays using different configurations. A promising approach to read out a large number of detectors with a SQUID array using a common transmission line is the microwave frequency domain read out. This type of read out can be leveraged to multiplex the signals from many detectors with high energy resolution and large signal bandwidth. A multiplexed microwave read out will be used to read out the HOLMES detectors [1], an array of 1000 micro-calorimeters based on dedicated TES (Transition Edge Sensors) [2], each coupled to a rf-SQUID. The HOLMES' detectors are Mo/Cu bilayer TESs with gold absorbers in which $^{163}$Ho ions will be implanted [3]. HOLMES aims at pushing the sensitivity on the neutrino mass below 1 eV by performing a calorimetric measurement of the energy released in the Electron Capture decay of $^{163}$Ho.

The rf-SQUID is a superconducting ring interrupted by a thin insulating layer, called weak link, through which Josephson tunnelling is possible. In a very simplified picture, by coupling the SQUID to a magnetic field the flux through the ring is quantized and the resulting current in the loop is [4], [5], [6], [7]:

$$i = I_o \sin\left(\frac{2\pi\Phi_{tot}}{\Phi_o}\right), \qquad \Phi_0 = \frac{h}{2e} \approx 2.07 \times 10^{-15} \text{ Wb} \qquad (1)$$

Where $I_0$ is a current dependent on SQUID/weak link composition, $\Phi_{tot}$ is the total induced flux and $\Phi_0$ is the flux quantum. Figure 1 shows a typical single channel rf-SQUID read out. The rf-SQUID is the circle with a cross, which represents the weak link. The rf-SQUID is inductively coupled to the TES, with the RF bias that sets the working point, $\Phi_{bias}$ (see the plot at the bottom of the figure) and the feedback, which forces the flux $\Phi_{feed}$ in the rf-SQUID opposite to the input flux $\Phi_{sig}$ in order to maintain the total flux constant, generating an unbalance in the amplifier output voltage proportional to $\Phi_{sig}$. The scheme of Figure 1 is over-simplified since the RF bias is removed from the signal by means of a not shown additional circuital block located at the amplifier output.

The classical set-up of Figure 1 is not convenient when one wants to minimize the connecting lines to the array, since every channel would need is own feedback link. To overcome this limitation the flux-ramp modulation, its simplified scheme is displayed in Figure 2, was recently successfully implemented [8], [9], [10]. Now the rf-SQUID is operated open loop and its bias is provided by the sawtooth signal generator ST. Provided that the period of the sawtooth is an integer multiple of the ratio of the flux quanta over the slope of the flux variation in time $d\Phi/dt$, the rf-SQUID current results at the bottom of the figure or (from (1)):

$$i = I_o \sin\left(\frac{2\pi(d\Phi/dt)t}{\Phi_o} + \frac{2\pi\Phi_{sig}}{\Phi_o}\right) \quad (2)$$

If the frequency of the sawtooth is much higher than the highest frequency component of the input signal, then the signal can be considered as a pure phase term. By evaluating the phase shift as a function of time, the signals due to particle interactions are properly reconstructed. This is depicted in Figure 2, where feedback is no more needed and the ST signal can be common to all the rf-SQUIDs of the array. Therefore, the ST signal is applied to a common transmission line composed of coils connected in series inductively coupled to every rf-SQUID (dashed link at ST in Figure 2). The output coil of the rf-SQUID is inductively coupled to a superconducting quarter wavelength resonator which, in turn, is coupled to a common feedline through a capacitance. Each resonator has different length, L1 … Ln in Figure 2, and the output signal from every rf-SQUID will modulate at a given, tuned frequency: the Sum_of_sines generator will therefore provide a superimposition of sinusoidal tones applied to this common line and read out by a cryogenic High Mobility Transistor (HEMT) amplifier. In this way, only two transmission lines are needed to operate the rf-SQUID array: one to read out the output signals and the other to provide the ramp signal [11].

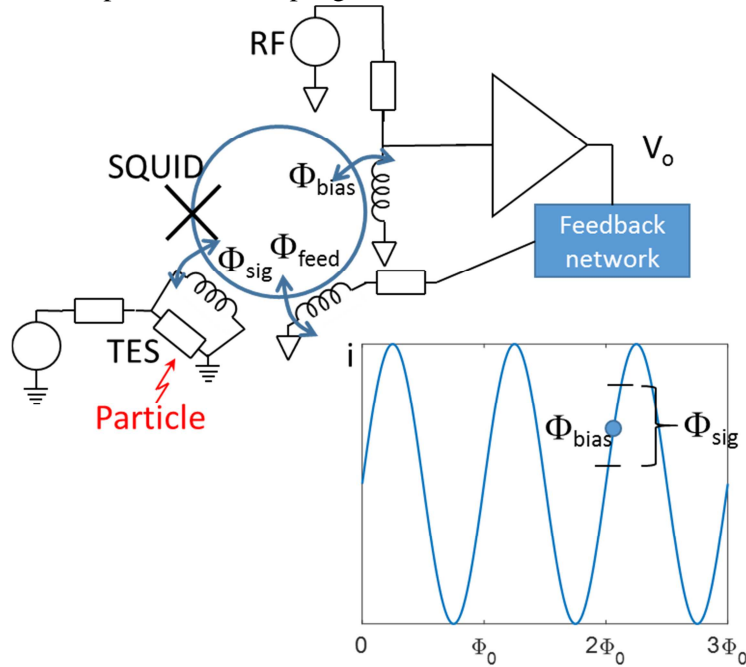

**Figure 1: rf-SQUID linked to a TES sensor and read out in closed loop configuration: $\Phi_{feed} \approx -\Phi_{sig}$ while the output voltage is proportional to $\Phi_{sig}$, by feedback action. The scheme is an over-simplification.**

As stated above, the sawtooth frequency must be larger than the highest frequency components of the input signal. At the same time, it is limited by the maximum number of ADC sample to take. For instance, for the final configuration of HOLMES detectors the ramp will be set to 500 kHz, as the detector signal rise time is expected close to 10 µs. Since the flux-ramp modulation has a crucial role for the reconstruction of the signal, any disturbances such as ground loop and EMI must be suppressed. In this paper, we describe our solution based on the use of a coupling transformer between the signal generator ST and the rf-SQUID coil of Figure 2. The transformer set-up covers an extended frequency range, starting from a few kHz (this latter property was necessary during the TES/SQUID initial studies). Two main requirements need to be satisfied. The rise of the sawtooth must be a straight line so that the oscillation pattern of the rf-SQUID is preserved over time; the fall time of the sawtooth must be fast enough to guarantee phase alignment.

We have developed a simple set up based on the use of commercial and inexpensive transformers for LAN networks, avoiding the design and construction of an ad hoc device. The set up allows to select a suitable number of transformers to be connected in series, so as to increase their low frequency input impedance, matching and preserving the sawtooth shape. In addition, the selection of the value of an impedance to be connected in parallel to the input coil of the transformer gives a bit of additional flexibility. In order to achieve a sufficiently clean sawtooth the lowest acceptable cut-off frequency of the transfer function of our set-up must be around ten times the frequency of the sawtooth ($f_{ramp}$), while the highest acceptable turn-on frequency must be lower than $f_{ramp}$. To confirm the validity of the developed system we characterized and modelled the transformers.

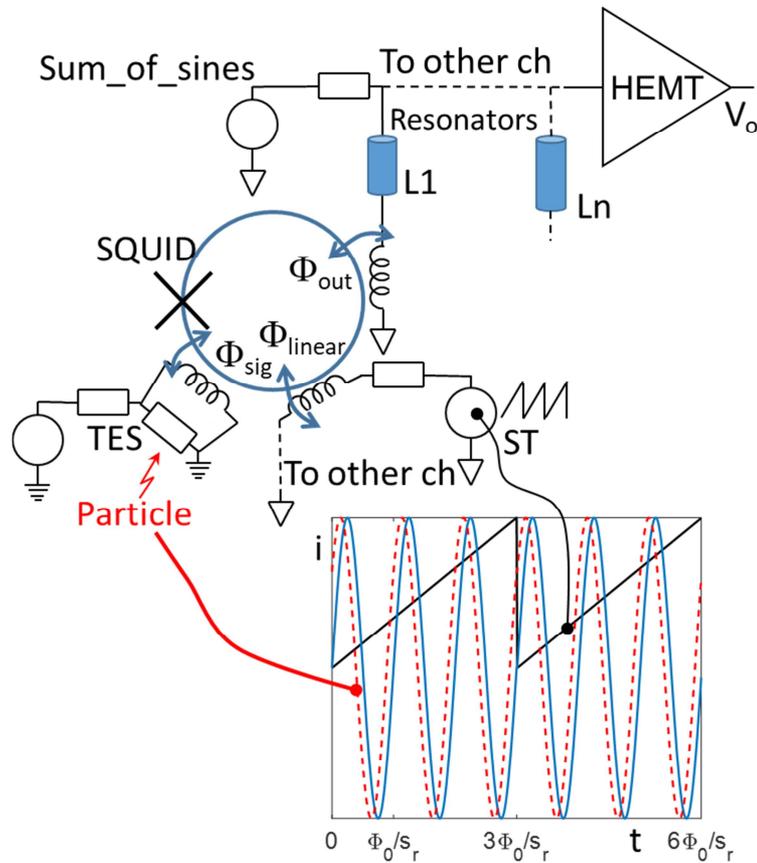

**Figure 2: rf-SQUID biased in open loop with the sawtooth flux-ramp modulation. If the sawtooth bias is faster than the TES signal, then the result is a phase-shift.**

## 2. Principle of operation and modelling

The idea of the setup of Figure 3 is simple: the source of the biasing signal drives the primary coil of the transformer through a coaxial line terminated at its input with $R_S$. The secondary coil of the transformer drives the load impedance $R_L$ at the end of the coaxial line. Primary and secondary coils of the transformer are equal and the ratio of the input to the output coil voltage is one. In this arrangement the input and output grounds are completely unconnected, and ground loops and EMI disturbances are strongly suppressed. To preserve the shape of the fast component of the sawtooth signal, reflection across the transmission lines must be avoided. Consequently, the terminating resistors $R_S$ and $R_L$ must be equal to the characteristic line value (50 Ω in our case), denoted as R in Figure 3. At high frequency, the impedance seen at the input coil equals the load impedance of the secondary coil in a 1:1 transformation ratio, so we do not expect signal reflection at the input.

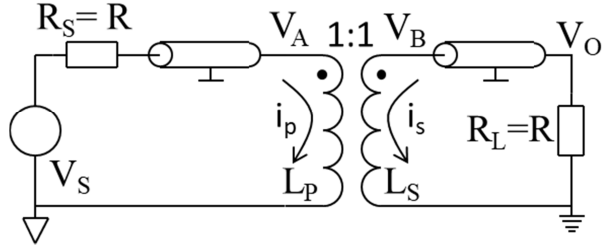

**Figure 3: Simplified set-up scheme.**

The system of equations that solve the network of Figure 3 must consider that the dropout voltage, $V_A$, across the input coil equals the sum of the voltage across the input inductance, $L_p$, and the voltage due to the mutual inductance, M; a similar consideration applies to $V_B$ and the output coil, $L_S$ (s is the complex frequency, s=$i\omega$):

$$V_A = sMi_s + sL_p i_p \qquad V_B = sMi_p + sL_s i_s \tag{3}$$

We can therefore write:

$$\begin{cases} \dfrac{V_S - V_A}{R} = \dfrac{V_A - sMi_s}{sL_p} \\ sMi_p + sL_s i_s + Ri_s = 0 \end{cases} \tag{4}$$

If we consider that $M = \sqrt{L_p L_s}$ and $L_s \approx L_p$ in our case, the solution of the system of equations (4) is:

$$V_O = -i_s R = \frac{sL_p}{2sL_p + R} V_i \tag{5}$$

The transfer function of the selected transformer, the surface mounted device WURTH ELEKTRONIK 749012011, was acquired with an Agilent AG4395A network-spectrum analyser. In Figure 4 measured data are compared with the model of Figure 3. From (5) the absolute value of the transfer function has a flat response at high frequencies, while the measured transfer function shows a high frequency roll-off. Therefore, the model of Figure 3 is not adequate. We increased a little the value of the load impedance $R_L$ noticing that the high frequency roll-off increased, proving the presence of a small parasitic inductance in series to the output. Then, we formulate the new model of Figure 5a that accounts for this inductance, $L_{PAR}$. The solving equations of the new model are:

$$\begin{cases} \dfrac{V_S - V_A}{R} = \dfrac{V_A - sMi_s}{sL_p} \\ sMi_p + sL_s i_s + sL_{PAR} i_s + Ri_s = 0 \end{cases} \quad (6)$$

To solve the system of (6) we use the MATLAB Symbolic Math toolbox and its least-squares solver. The numerical values of all parameters are obtained by fitting the data. As shown in Figure 4 the model (dash-dot line) is still not adequate. At high frequency, the roll-off is accounted by the presence of the parasitic inductance $L_{PAR}$, but, due to the presence of parasitic capacitances in parallel to the coil, the high frequency roll-off is less steep than that expected from (6). In order to take into account the high frequency behaviour of the transformer, we made the simple assumption that the inductances of the input and output windings are bonded by a frequency-dependent proportionality factor. Therefore, we substituted the ratio 1:1 of Figure 5a with the ratio 1: $a(\omega)$ as shown in Figure 5b. As already assumed in (6), we considered an ideal transformer with $M = \sqrt{L_p L_s}$ and we added $L_{PAR}$. In addition to the previous model, the input and output coil inductances, $L_P$ and $L_S$, now differ only in frequency behaviour. We also multiplied the output voltage by a coefficient, K, that accounts for tolerance in the impedance values and the link; the coefficient is found to be close to one.

Since the high frequency roll-off is dependent on $R_L/L_{PAR}$, we considered a polynomial ratio with equal number of zeros and poles to describe the relation between $L_P$ and $L_S$:

$$L_s = a(\omega)^2 L_p = \left( \frac{(1 + sZ_A)(1 + sZ_B)}{(1 + sP_A)(1 + sP_B)} \right)^2 L_p \quad (7)$$

The squaring of $a(\omega)$ is convenient as the final transfer function is linearly dependent on $a(\omega)$, while in the denominator its square is present in a term that comes into play only at high frequencies, where $|a(\omega)|$ is close to one. We forced the poles to have a real negative part to ensure stability, but we left the real part of the zeros to be negative or positive to shape the phase properly. The minimum number of zeros (and poles) for a more than adequate result is two. From the fit, all the poles and the zeros are found real. These fitting parameters take into account the parasitic capacitances between the coils of $L_p$ and $L_s$.

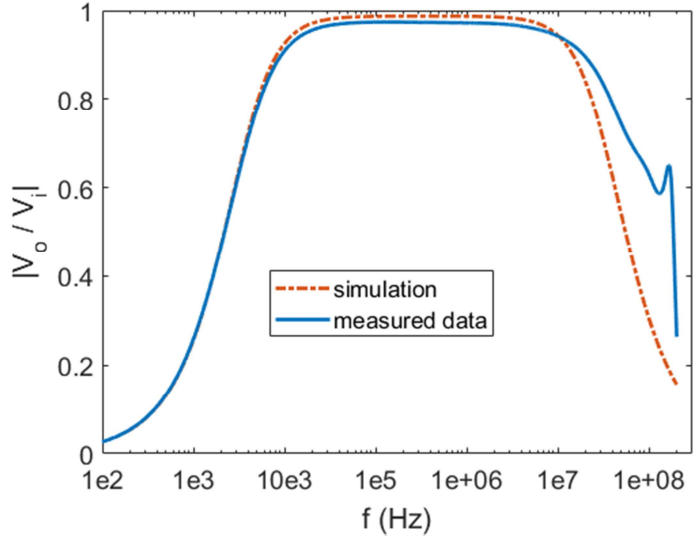

**Figure 4: Module of the transfer function of Figure 3 and equations (4), measured with the Agilent AG4395A, continuous line, simulated with the model of (Figure 5), dash-dot line. Measured data are downloaded from the network-spectrum analyser.**

The new model of (6) and (7) was used to fit the measured data of Figure 4, shown now in Figure 6 with both the real and imaginary components superimposed. The model fits the measured data fairly well, except for a small deviation at high frequency. The proof of the accuracy of the model is even more evident in Figure 7, where the step response of the transformer is shown together with the modelled response, obtained using the parameters extracted from the fit in Figure 6. It is interesting to observe that the model works with no need to include (parasitic) capacitances, which are in some way embedded in (7).

The ratio $V_o/V_s$, obtained by (6) and (7), of the circuit of Figure 5b is a long expression, not reported here. The simplified model of the circuit shown in Figure 5b is displayed in Figure 8 where we consider an ideal transformer, a quadrupole with ratio $a(\omega)$ between its output and input, whose input impedance is $sL_{PAR}+R_L$ in parallel to $L_P$. Furthermore, we add resistor $R_{IN}$ and $R_{SER}$ for a reason that will be explained below. The network has now independent input and output loops, and is easily solved:

$$\begin{cases} V_A = \dfrac{sL_P \| R_{IN} \| (R_L + sL_{PAR})}{sL_P \| R_{IN} \| (R_L + sL_{PAR}) + R + R_{SER}} V_S \\ \\ V_O = a(\omega) \dfrac{R_L}{sL_{PAR} + R_L} V_A, \qquad a(\omega) = \dfrac{(1 + sZ_A)(1 + sZ_B)}{(1 + sP_A)(1 + sP_B)} \end{cases} \qquad (8)$$

Now we have to analyse the behaviour of the transformer response to a sawtooth input signal. Figure 9 shows this for a 5 kHz frequency that highlights the limitation: the linear rising part is largely bent. The reason for this is understood looking at the first relation of (8): $V_A$ is AC coupled to $V_S$, with a zero in the origin. The proportionality becomes flat only above the frequencies at which $sL_P$ approaches $R_L\|R_{IN}$. To lower this frequency we can follow 2 methods: 1) to lower the value of $R_L\|R_{IN}$: in this case signal $V_A$ will be attenuated and $V_S$ will need to be increased to maintain the appropriate output value (there is a lower limit in frequency beyond which the transformer core saturates); 2) to increase the value of the inductance $L_p$ of the primary coil. We used both approaches, as described below.

The approach 2) is often pursued by designing an ad hoc transformer with optimized number of turns. With a large number of turns, the impact of parasitic capacitance and a possible increase in $L_{PAR}$ should be considered. Since we need to face a broad range of sawtooth signals from a few kHz up to a MHz, we chose to implement a set-up that we can call a trimmable-transformer, consisting of several transformers, as the one characterized above, connected in series in order to match the small/high frequency needs.

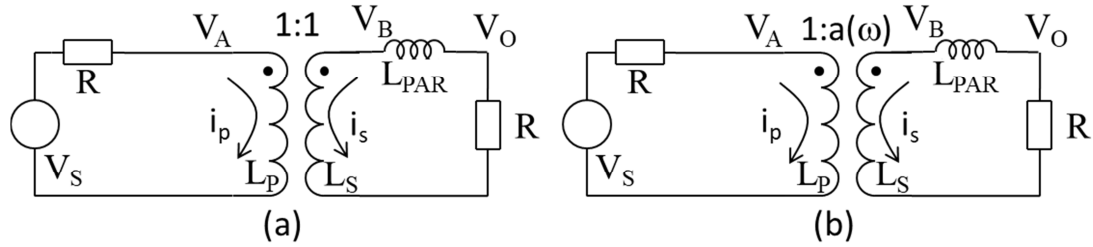

**Figure 5: Circuit model of Figure 3 improved by $L_{PAR}$ and (a) constant ratio of voltage between secondary and primary and (b) frequency dependent ratio, a(ω).**

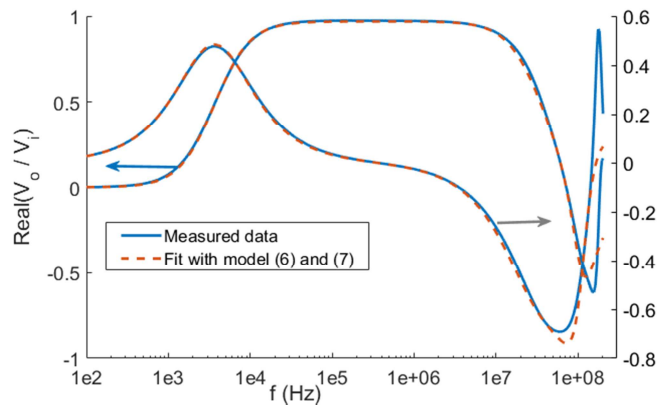

**Figure 6: Measured and fitted real and imaginary components of the transfer function modelled with (6) and (7). The continuous lines are measured data, the dash-dot lines are the fittings. Measured data are downloaded from the network-spectrum analyser.**

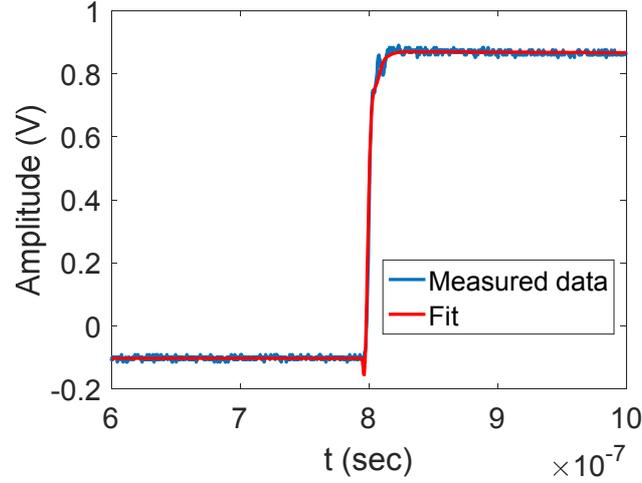

**Figure 7: Transformer response to a step voltage input (csv file from the scope) superimposed to the simulation, red line, from the circuit model of Figure 5, (6) and (7). Signals are normalized in amplitude. Rise time is less than 8 ns.**

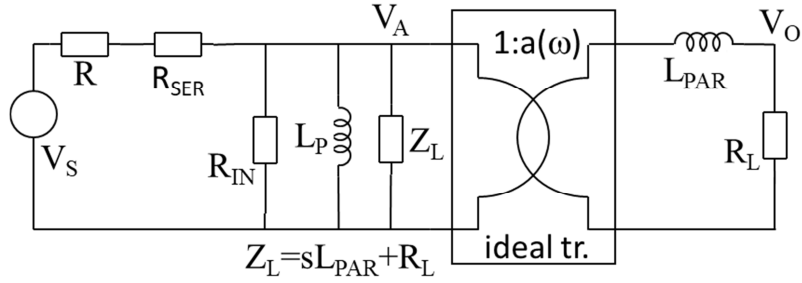

**Figure 8: Simplified model of the circuit of Figure 5.**

## 3. The trimmable-transformer set-up and results

### 3.1 Circuit description

The circuit arrangement of our trimmable-transformer is shown in Figure 10. A number *n* of transformers are connectable in series thanks to the presence of a set of switches pairs SWxa and SWxb, 1≤x≤*n*, which, operated in the same way, allow to enable (open position) or not (short position) the corresponding transformer. In Figure 10, to simplify, we consider the circuit model of Figure 8 for every transformer, represented by its input coil, $L_p$, output parasitic inductance, $L_{PAR}$, and transfer ratio, $a(\omega)$. When the switches are in the open state, the primary has at its input the load impedance $Z_L$ divided by *n*, as shown. If *k* is the number of switches left open, then the equivalent circuit looks as in Figure 11 and:

$$\begin{cases} V_A = \dfrac{skL_P \| R_{IN} \| (R_L + skL_{PAR})}{skL_P \| R_{IN} \| (R_L + skL_{PAR}) + R} V_S \\[2ex] V_O = a(\omega)\dfrac{R_L}{skL_{PAR} + R_L} V_A, \quad a(\omega) = a\dfrac{(1 + sZ_A)(1 + sZ_B)}{(1 + sP_A)(1 + sP_B)} \end{cases} \quad (9)$$

We can appreciate in the first equation of (9) that now the primary coil has an impedance $k$ time larger, or $skL_p$, and, since $R_{IN}$ and $R_L$ remain the same, the low frequency coupling is shifted at low frequency. At the same time from the second equation of (9) the parasitic impedance at the secondary coil is increased to $skL_{PAR}$ and the high frequency roll-off is reduced, too. Increasing the number of operated transformers both the low frequency coupling and the high frequency roll-off are lowered, so a trade-off can be found, as the lower is the sawtooth frequency, the less is demanded to the speed of its fast transition, and vice versa for a large frequency sawtooth.

The WURTH ELEKTRONIK 749012011 has a pair of transformers. We used 3 of such devices for a total of 6 transformers coupled, each one, with a pair of configurable switches. The photograph of Figure 12 shows our layout.

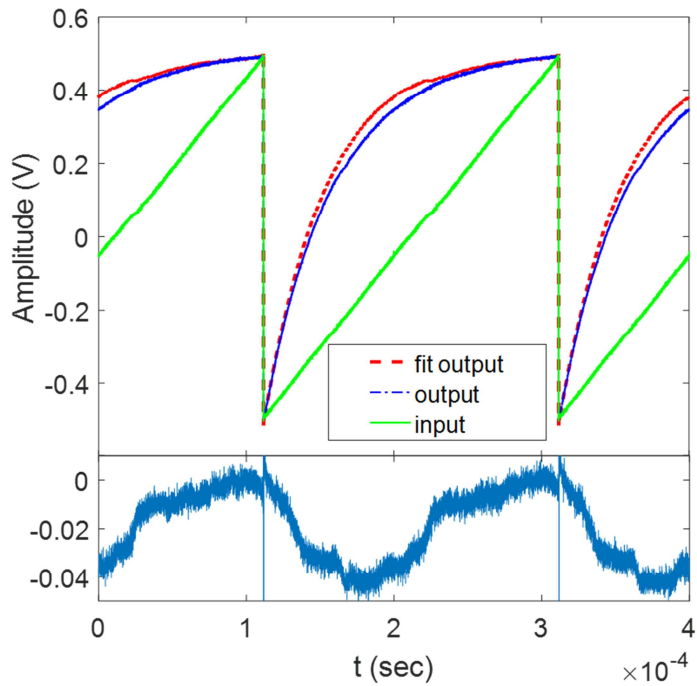

**Figure 9: Transformer answer to a 5 kHz sawtooth signal. The continuous line is the input signal, the red dash-dot and the blue lines are the fitted and measured signals. The agreement between data and the model is at the level of 4%. Signals are normalized in amplitude. Input signal and measured data are downloaded from the scope.**

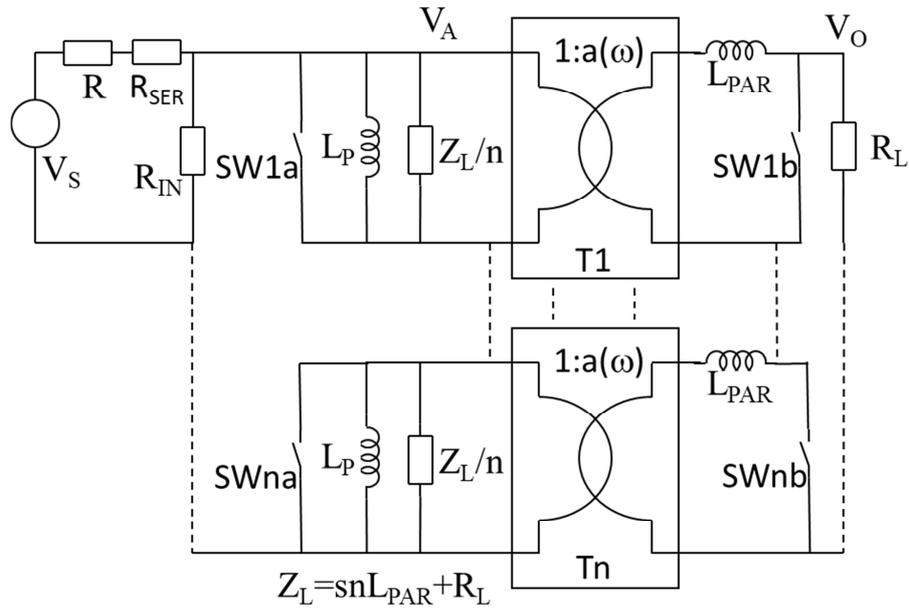

**Figure 10: Series arrangement of the transformer of Figure 8. Impedance values above are for the case the switches are all open.**

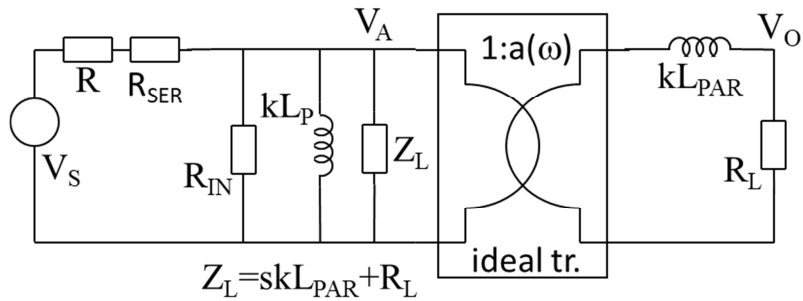

**Figure 11: Model of the set-up of Figure 10.**

## 3.2 Measurement results

Figure 13 shows the response of our trimmable-transformer, superimposed to a 5 kHz sawtooth input signal when 6 transformers are connected in series, $R_{IN}$=4.7 Ω and $R_{SER}$=47 Ω. $R_{SER}$ is at the receiving end of the coaxial cable. It is used to set the input impedance of the circuit, given by the sum of $R_{SER}$ itself and the impedance at the transformer input, at 50 Ω. We can see now that the response matches the input, compared to the response of Figure 9. The response to a faster sawtooth of 1 MHz frequency is shown in Figure 14, where only one transformer is used with $R_{IN}$=∞ and $R_{SER}$=0. In this way the fast transition is satisfied, as well. Another example is reported in Figure 15 where the signal is 10 kHz, $R_{IN}$=11 Ω ($R_{SER}$=39 Ω) and k=4: the output signal is almost indistinguishable from the input signal.

Fitting curves of Figure 13, Figure 14 and Figure 15 have been obtained by extracting the parameters from the measurements taken with the spectrum analyser. The accuracy of the injected signal and scope data is at about 1% level and this limits our residuals evaluation. In Figure 16 we emulated the sine of (2) at 25 kHz, when the sawtooth is at 5 kHz, for the case $R_{IN}$=4.7 Ω and $R_{SER}$=47 Ω. This sawtooth frequency is the lowest that ensures a good generation with a negligible error with respect to the sinusoid generated from the input signal.

Simulated and measured transfer functions of the series of one to 6 transformers are shown in Figure 17 and Figure 18 for the real and imaginary parts. The fitting function used in this case is that of (8). Inductance $L_P$ is about 1090 µH, which grows linearly up to about 6200 µH when 6 transformer are put in series. The parasitic inductor $L_{PAR}$ was also found linearly dependent on the number of the transformers with a value that is close to 500 nH, 1 transformer, up to 1700 nH, 6 transformers. A pedestal of about 200 nH results in this latter case, probably coming from our board layout.

The sensitivity to $Z_A$, $Z_B$, $P_A$ and $P_B$ lowers with the number of transformers as the roll-off due to $L_P$ and $R_L$ dominates. With one transformer it resulted -72 MHz, 138 MHz, for $Z_A$ and $Z_B$, respectively, and around -150 MHz, for $P_A$ and $P_B$. A summary of the fit parameters is in Table 1.

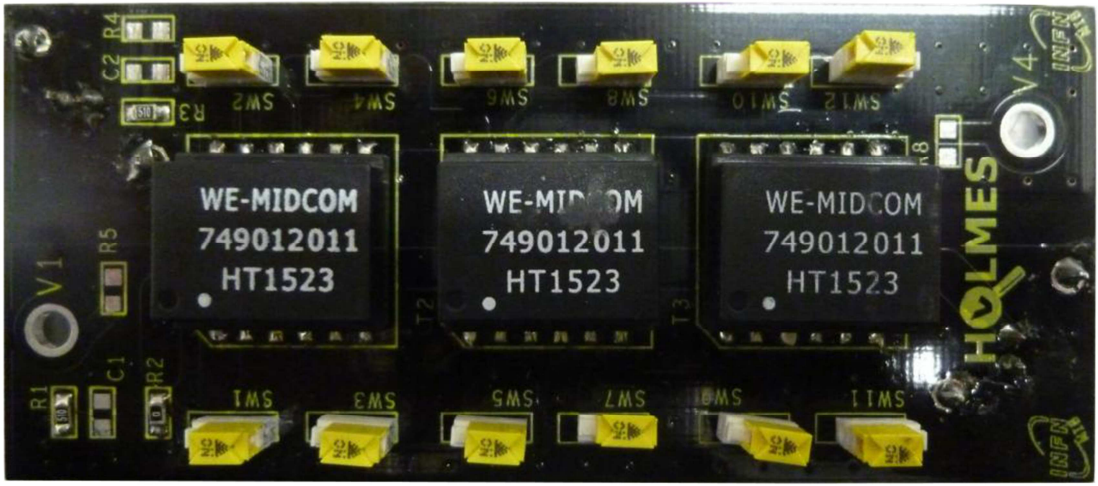

**Figure 12: Photograph of the circuit set-up. The 6 transformers are contained in the 3 SMD circuits in the center of the boards. The 6 pairs of yellow switches are SW1, ..., SW12.**

**Table 1: Summary of the parameters extracted from the measured data with our model.**

|  | $L_P$ (µH) | $L_{PAR}$ (nH) | K | $Z_A$ (MHz) | $Z_B$ (MHz) | $P_A$ (MHz) | $P_B$ (MHz) |
|---|---|---|---|---|---|---|---|
| Pedestal | 0 | 200 | 0.97 |  |  |  |  |
| Slope | 1030 | 300 | -0.008 |  |  |  |  |
| Valid for n=1, only |  |  |  | -72 | 138 | -150 | -150 |

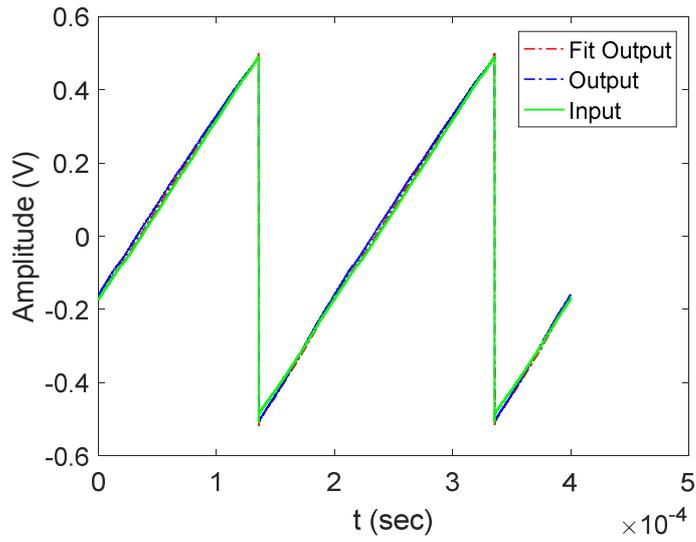

**Figure 13:** Measured and fitted sawtooth response with 6 transformers connected in series, k=6, and $R_{IN}$=4.7 Ω, superimposed to the 5 kHz input signal. Fitted and measured signals are indistinguishable. Signals are normalized in amplitude. Input signal and measured data are downloaded from the scope, csv files.

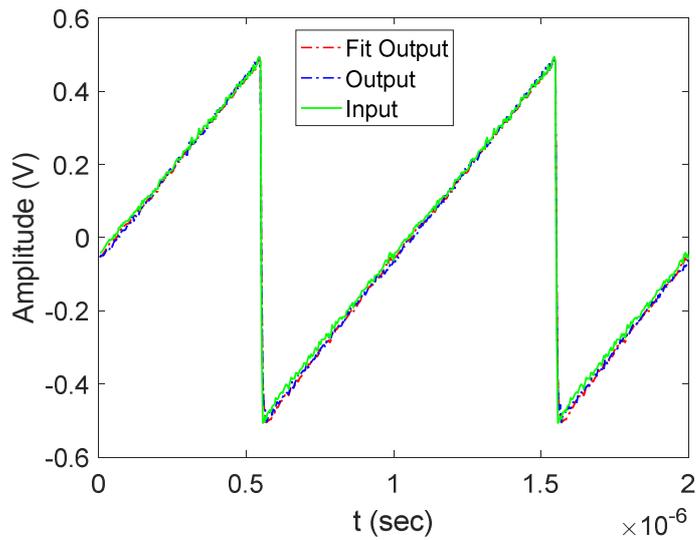

**Figure 14:** Measured and fitted sawtooth response with only one transformer, k=1, superimposed to the 1 MHz input signal. Fitted and measured signals are indistinguishable. Signals are normalized in amplitude. Input signal and measured data are downloaded from the scope, csv files.

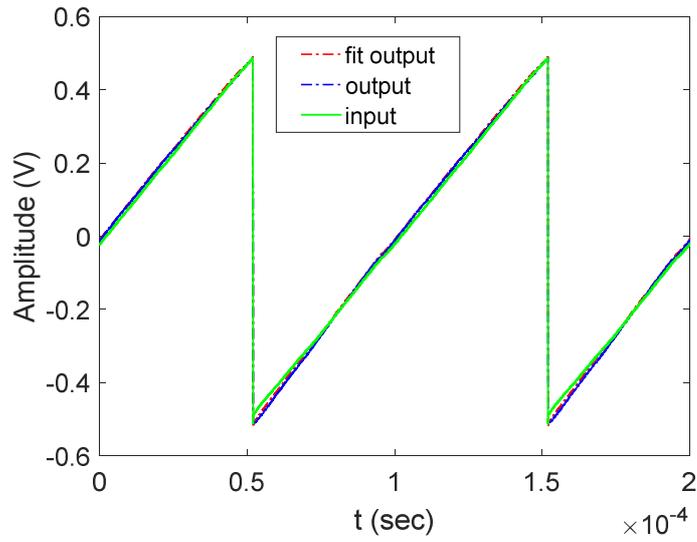

**Figure 15**: Measured and fitted sawtooth response with four transformers, k=4 and $R_{IN}$=11 Ω, superimposed to 10 kHz input signal. Fitted and measured signals are indistinguishable. Signals are normalized in amplitude. Input signal and measured data are downloaded from the scope, csv files.

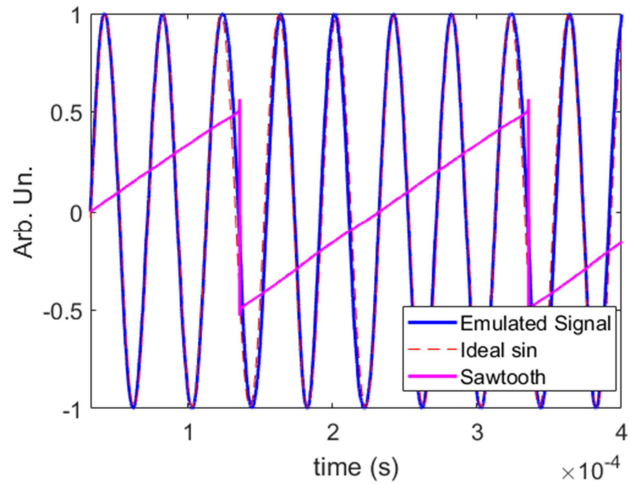

**Figure 16**: Emulated sinusoid of (2) when the sawtooth is 5 kHz, $R_{IN}$=4.7 Ω, the generated sinusoid is at 25 kHz and k=6. This is the minimum frequency and number of transformers to obtain a signal similar to that obtained from the input sawtooth.

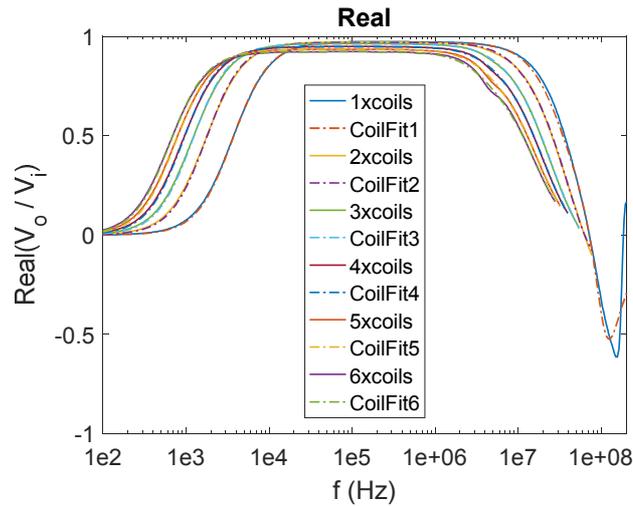

**Figure 17: Fits, dash-dot lines, and measured data, continuous lines, of the real part of the transfer function of the circuit configuration of Figure 10 when the number of transformers connected in series varies from 1 to 6.**

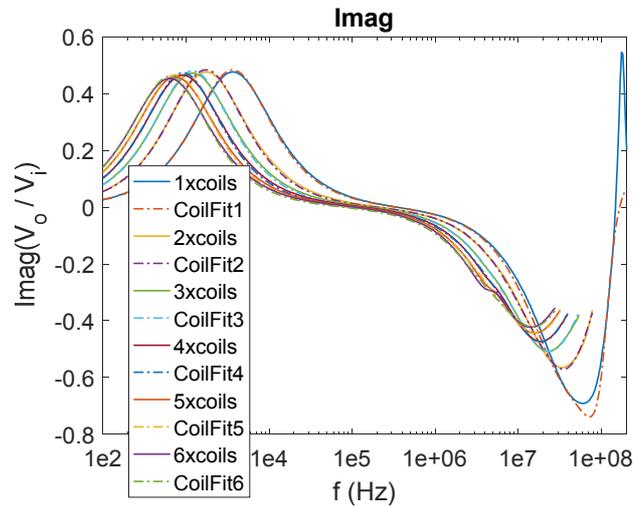

**Figure 18: Fits, dash-dot lines, and measured data, continuous lines, of the imaginary part of the transfer function of the circuit configuration of Figure 10 when the number of transformers connected in series varies from 1 to 6.**

## 4. Conclusion

The necessity of an AC bias for rf-SQUIDs which does not suffer from ground loop disturbances and EMI has triggered the developing and modelling of the trimmable-transformer, a circuit set-up which enables to implement a 1:1 transformer ratio with different values of the primary and secondary coils inductances. The trimmable-transformer is composed of commercial SMD standard transformers, and it can be configured to match the required AC coupling frequency with adequate bandwidth. Changing the number of transformers from 1 to 6 or $R_{SER}$ value results in a pass-band shift by a factor ~14. This allows to cover sawtooth signals over the frequency range of our application, from 5 kHz to 1 MHz.

## 5. Acknowledgements

This work was supported by the European Research Council (FP7/2007-2013) under Grant Agreement HOLMES no. 340321.